\documentclass[aps,prl,amsmath,amssymb,twocolumn,superscriptaddress]{revtex4}
\usepackage{graphicx}
\usepackage{braket}
\usepackage{color}
\usepackage{soul}

\begin{document}


\title{Notch filtering the nuclear environment of a spin qubit}

\author{Filip~K.~Malinowski}
\thanks{These authors contributed equally to this work}
\affiliation{Center for Quantum Devices and Station Q Copenhagen, Niels Bohr Institute, University of Copenhagen, 2100 Copenhagen, Denmark}

\author{Frederico~Martins}
\thanks{These authors contributed equally to this work}
\affiliation{Center for Quantum Devices and Station Q Copenhagen, Niels Bohr Institute, University of Copenhagen, 2100 Copenhagen, Denmark}

\author{Peter~D.~Nissen}
\affiliation{Center for Quantum Devices and Station Q Copenhagen, Niels Bohr Institute, University of Copenhagen, 2100 Copenhagen, Denmark}

\author{Edwin~Barnes}
\affiliation{Department of Physics, Virginia Tech, Blacksburg, Virginia 24061, USA}
\affiliation{Condensed Matter Theory Center and Joint Quantum Institute, Department of Physics, University of Maryland, College Park, Maryland 20742-4111, USA}

\author{\L ukasz~Cywi\'nski}
\affiliation{Institute of Physics, Polish Academy of Sciences, al.~Lotnik{\'o}w 32/46, PL 02-668 Warsaw, Poland}

\author{Mark~S.~Rudner}
\affiliation{Center for Quantum Devices and Station Q Copenhagen, Niels Bohr Institute, University of Copenhagen, 2100 Copenhagen, Denmark}
\affiliation{Niels Bohr International Academy, Niels Bohr Institute, 2100 Copenhagen, Denmark}

\author{Saeed~Fallahi}
\affiliation{Department of Physics and Astronomy, Birck Nanotechnology Center, and Station Q Purdue, Purdue University, West Lafayette, Indiana 47907, USA}

\author{Geoffrey~C.~Gardner}
\affiliation{Department of Physics and Astronomy, Birck Nanotechnology Center, and Station Q Purdue, Purdue University, West Lafayette, Indiana 47907, USA}
\affiliation{School of Materials Engineering, Purdue University, West Lafayette, Indiana 47907, USA}

\author{Michael~J.~Manfra}
\affiliation{Department of Physics and Astronomy, Birck Nanotechnology Center, and Station Q Purdue, Purdue University, West Lafayette, Indiana 47907, USA}
\affiliation{School of Materials Engineering, Purdue University, West Lafayette, Indiana 47907, USA}

\author{Charles~M.~Marcus}
\affiliation{Center for Quantum Devices and Station Q Copenhagen, Niels Bohr Institute, University of Copenhagen, 2100 Copenhagen, Denmark}

\author{Ferdinand~Kuemmeth}
\affiliation{Center for Quantum Devices and Station Q Copenhagen, Niels Bohr Institute, University of Copenhagen, 2100 Copenhagen, Denmark}

\newcommand{\VL}{V_\mathrm{L}}
\newcommand{\VM}{V_\mathrm{M}}
\newcommand{\VR}{V_\mathrm{R}}

\newcommand{\Btot}{B^\mathrm{tot}}
\newcommand{\Bext}{B^\mathrm{ext}}
\newcommand{\Bznuc}{B_\mathrm{z}^\mathrm{nuc}}
\newcommand{\Bpnuc}{B_\perp^\mathrm{nuc}}

\newcommand{\ud}{\uparrow\downarrow}
\newcommand{\du}{\downarrow\uparrow}

\newcommand{\drv}{\mathrm{d}}

\newcommand{\FHahn}{F_\mathrm{Hahn}}
\newcommand{\FCPMG}{F_{\mathrm{CPMG},n}}
\newcommand{\FFID}{F_\mathrm{FID}}
\newcommand{\Fenv}{F_\mathrm{env}}

\newcommand{\Ga}{^{69}\mathrm{Ga}}
\newcommand{\Gb}{^{71}\mathrm{Ga}}
\newcommand{\As}{^{75}\mathrm{As}}
\newcommand{\fGa}{f_{^{69}{\rm Ga}}}
\newcommand{\fGb}{f_{^{71}{\rm Ga}}}
\newcommand{\fAs}{f_{^{75}{\rm As}}}

\newcommand{\TCPMG}{T_2 ^\mathrm{CPMG}}
\renewcommand{\vec}[1]{{\bf #1}}

\date{\today}

\begin{abstract}

\end{abstract}

\maketitle

\textbf{
Electron spins in gate-defined quantum dots provide a promising platform for quantum computation \cite{Loss1998,Veldhorst2015,Nowack2011,Bluhm2011,Petta2005,Nowack2007,Maune2012}.
In particular, spin-based quantum computing in gallium arsenide takes advantage of the high quality of semiconducting materials, reliability in fabricating arrays of quantum dots, and accurate qubit operations \cite{Nowack2007,Petta2005,Foletti2009,Dial2013,Maune2012,Martins2016}.
However, the effective magnetic noise arising from the hyperfine interaction with uncontrolled nuclear spins in the host lattice constitutes a major source of decoherence \cite{Petta2005,Bluhm2011,Botzem2016,Martins2016}. 
Low frequency nuclear noise, responsible for fast (10 ns) inhomogeneous dephasing \cite{Petta2005}, can be removed by echo techniques \cite{Petta2005,Bluhm2011,Viola1998,Barthel2010,Medford2012,Botzem2016}. High frequency nuclear noise, recently studied via echo revivals \cite{Bluhm2011,Botzem2016}, occurs in narrow frequency bands related to differences in Larmor precession of the three isotopes $\mathbf{^{69}Ga}$, $\mathbf{^{71}Ga}$,  and $\mathbf{^{75}As}$ \cite{Cywinski2009,Cywinski2009a, Neder2011}.
Here we show that both low and high frequency nuclear noise can be filtered by appropriate dynamical decoupling sequences, resulting in a substantial enhancement of spin qubit coherence times. Using nuclear notch filtering, we demonstrate a spin coherence time ($\mathbf{T_{2}}$) of 0.87 ms, five orders of magnitude longer than typical exchange gate times, and exceeding the longest coherence times reported to date in Si/SiGe gate-defined quantum dots~\cite{Eng2015,Kawakami2016}.
}

\begin{figure}
	\includegraphics[scale=1]{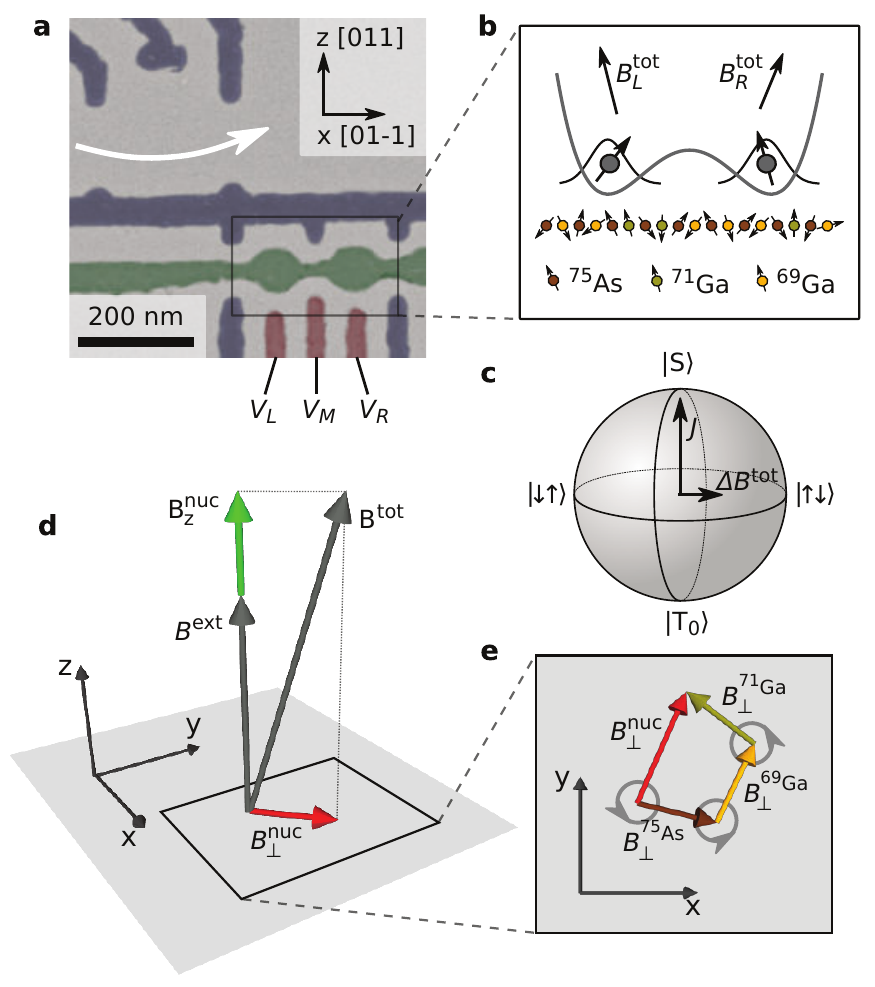}
	\caption{\textbf{Singlet-triplet qubit interacting with a nuclear spin bath.}
	\textbf{a}, False-color scanning electron micrograph of a device similar to the one measured, consisting of a double dot (surrounded by black rectangle) and a proximal readout dot (indicated by white arrow).
	\textbf{b}, Double-well potential occupied by two electrons. Within the left (right) dot an effective magnetic field $\Btot_\mathrm{L(R)}$ splits the electron spin states due to the Zeeman effect and hyperfine interaction with spinful nuclei of $^{69}$Ga, $^{71}$Ga, and $^{75}$As.
	\textbf{c}, Bloch sphere representation of the qubit with corresponding two-electron spin states indicated. Two rotation axes are defined by the exchange interaction, $J$, and 
the total field gradient between the dots, $\Delta \Btot=\Btot_\mathrm{L}-\Btot_\mathrm{R}$.
	\textbf{d}, The effective magnetic field $\Btot$ acting on each spin is set by 
the external magnetic field $\vec{B}^{\rm ext}$ (nominally aligned with the [011] crystal axis), the slowly fluctuating Overhauser field component $\vec{B}^\mathrm{nuc}$ parallel to $\vec{\Bext}$, and the rapidly changing transverse Overhauser field $\vec{B}_\perp^\mathrm{nuc}$. 
Here we suppress the dot label indices for brevity.
	\textbf{e}, The transverse Overhauser field $\vec{B}^{\rm nuc}_\perp = \vec{B}_\perp^{^{69}\mathrm{Ga}} + \vec{B}_\perp^{^{71}\mathrm{Ga}} + \vec{B}_\perp^{^{75}\mathrm{As}}$ is a sum of fields
of the three nuclear species, each precessing at its Larmor frequency.
	}
	\label{fig1}
\end{figure}

The qubit under study is implemented in a gate-defined double dot, with a potential that can be manipulated via nanosecond voltage pulses applied to gate electrodes $\VL$, $\VM$ and $\VR$ (Fig.~\ref{fig1}a and Methods). 
The qubit states are encoded in the two-electron spin singlet state, $\ket{S} = \frac{1}{\sqrt{2}}(\ket{\ud} - \ket{\du})$, and the spin triplet state,  $\ket{T_0} = \frac{1}{\sqrt{2}}(\ket{\ud} + \ket{\du})$, where the arrows indicate the spin projections of the electrons in the left and right dots \cite{Petta2005,Foletti2009}. 
These qubit states are energetically separated from the spin-polarized two-electron states, $\ket{\uparrow\uparrow}$ and $\ket{\downarrow\downarrow}$, by an external magnetic field $\vec{B}^{\rm ext}$, ranging from 0.2 to 1 tesla in this experiment.
Single-shot readout of the qubit is accomplished using spin-to-charge conversion followed by readout of a proximal sensor dot \cite{Barthel2009, Barthel2010} (see Methods).

As illustrated in Fig.~\ref{fig1}b,d, 
the local Zeeman energy in dot $d = {\rm L,R}$ is perturbed by the Overhauser field $\vec{B}^{{\rm nuc}}_d$ arising from the hyperfine interaction with the nuclear spin bath. 
In our device, each electron is in contact with $\sim 10^6$ nuclear spins, comprised of three species:
 $^{69}$Ga, $^{71}$Ga, and $^{75}$As \cite{Petta2005,Bluhm2011,Neder2011}. 

The Bloch sphere of the $S$-$T_0$ qubit is shown in Fig.~\ref{fig1}c.  Bold arrows indicate the rotation axes associated with the exchange interaction, $J$, and the gradient of the effective 
field between the dots, $\Delta \Btot = \Btot_{\rm L} - \Btot_{\rm R}$, where $\Btot_d = \sqrt{|\vec{B}^{\rm ext} + \vec{B}^{{\rm nuc}}_d|^2}$ is the magnitude of the total effective field in dot ${d}$~\cite{Foletti2009}.
Note that transverse nuclear field gradients tilt the quantization axes in the two dots relative to each other.
For large external fields this primarily leads to a minor redefinition of the qubit subspace~\cite{Neder2011}; for simplicity throughout this work we refer to the states in the qubit subspace by the conventional labels $S$ and $T_0$.

Overhauser field fluctuations in each dot are non-Markovian, with low frequency (power-law) spectral content parallel to the external field, denoted $\Bznuc$ (suppressing the dot index), and narrow-band spectral components at the nuclear Larmor frequency scale perpendicular to the external field, denoted $\vec{B}^{\rm nuc}_\perp$. Low frequency fluctuations arise primarily from nuclear spin diffusion~\cite{Reilly2008}, driven by dipole-dipole interactions between neighboring nuclei, and nonlocal electron-mediated flip-flops \cite{DeSousa2003,Yao2006,Cywinski2009,Cywinski2009a}.
High frequency fluctuations of $\vec{B}^{\rm nuc}_\perp$ arise primarily due to the megahertz-scale relative Larmor precession of different nuclear spins~\cite{Bluhm2011,Neder2011,Botzem2016}. 
The transverse Overhauser field $\vec{B}^{\rm nuc}_\perp$ is given by the sum of contributions $\vec{B}_\perp^{^{69}\mathrm{Ga}}$, $\vec{B}_\perp^{^{71}\mathrm{Ga}}$, and $\vec{B}_\perp^{^{75}\mathrm{As}}$ of the three isotopic species, each of which precesses at its own  Larmor frequency, see Fig.~\ref{fig1}e.
This leads to modulations of the total field in each dot, $\Btot$, which are concentrated near the {\it differences} of the nuclear Larmor frequencies, and contribute quadratically to the qubit splitting.
 

\begin{figure*}
	\includegraphics[scale=1]{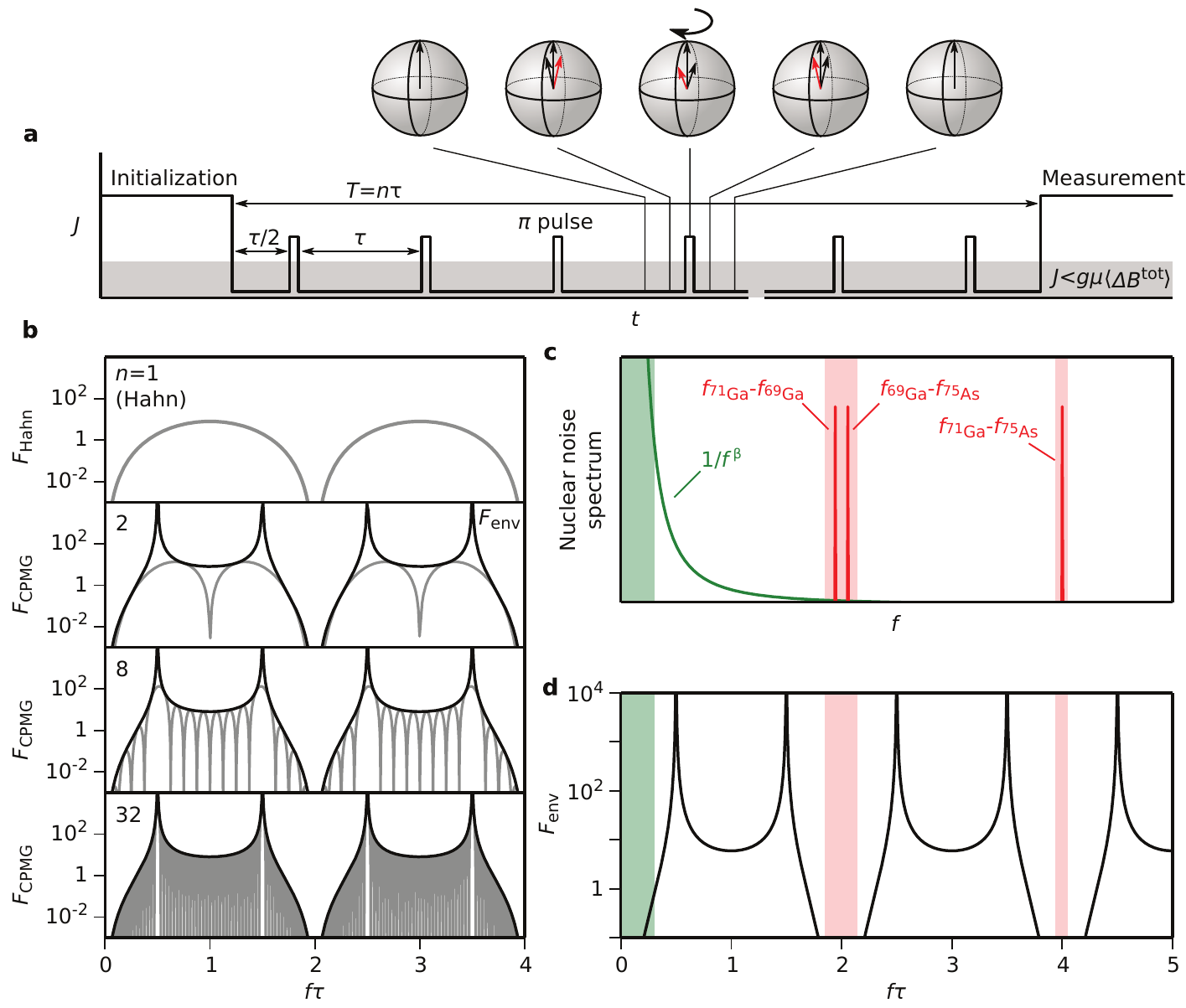}
	\caption{\textbf{Frequency-selective dynamical decoupling.}
	\textbf{a}, CPMG pulse sequence consisting of $n$ pulses separated by time $\tau$. At the beginning, two electrons prepared in a singlet state $|S\rangle$ (Initialization) are rapidly separated into two dots with negligible exchange splitting (shaded region of $J$). After a total separation time $T=n \tau$ the preserved qubit state is detected by the readout dot via spin-to-charge conversion (Measurement). During the separation time the two-electron state evolves in the fluctuating gradient of total magnetic field $\Delta \Btot$. For slow fluctuations, the phases acquired before and after each $\pi$ pulse cancel each other, due to the sign reversal of the acquired phase by the $\pi$ pulse.
This is exemplified for three different values of $\Delta \Btot$ by arrows in the Bloch sphere. 
	\textbf{b}, Filter functions of Hahn echo ($n$ = 1) and CPMG sequence with $n$ = 2, 8 and 32 $\pi$ pulses (gray). Envelope of the filter function $\Fenv$ reveals a frequency selectivity that is independent of $n$ (black).
	\textbf{c}, Schematic spectral density of nuclear noise. The linear low frequency part (green), described by a power law, is dominated by fluctuations associated with diffusion of the longitudinal component of the nuclear spin. The quadratic high frequency noise (red) results from fluctuations of $\Btot$ at differences of nuclear Larmor frequencies.
	\textbf{d}, By adjusting the time between $\pi$ pulses, the minima of the filter function envelope $\Fenv$ (black) can be aligned with the nuclear noise spectrum (green and red shading), thereby decoupling the qubit from both linear low frequency and quadratic high frequency noise. 
	}
	\label{fig2}
\end{figure*}
To decouple the qubit from the multiscale nuclear noise, we employ the Carr-Purcell-Meiboom-Gill (CPMG) pulse sequence shown in Fig.~2a. 
We first initialize the double dot in a spin singlet by temporarily loading two electrons into the left dot.
Then we quickly separate the electrons in the double-well potential, thereby rapidly turning off the exchange interaction, $J$. In this configuration, the gradient of the total effective field, $\Delta \Btot$, causes uncontrolled qubit rotation around the horizontal axis of the Bloch sphere in Fig.~\ref{fig1}c. 
After a time $\tau/2$, an exchange pulse is applied by temporarily lowering the barrier between dots with a voltage pulse on gate $\VM$ \cite{Martins2016}, implementing a $\pi$ rotation around the vertical axis of the Bloch sphere (see Supplementary Section 1).
We repeat this set of operations $n$ times (where $n$ is even) and, after a total evolution time $T = n\tau$, read out the state of the qubit. The fraction of singlet outcomes is denoted $P_S$. Setting $n=1$ implements a Hahn-echo sequence, and allows comparison to previous work \cite{Bluhm2011,Childress2006,Botzem2016}.

For quasistatic nuclear noise, the effective field acting on the qubit before and after the $\pi$ pulse is nearly the same, causing the qubit state to be refocused to the singlet after an interval $\tau/2$.
For nuclear noise with power spectrum $S(f)$, Hahn and CPMG sequences 
act as a filter of the noise in the frequency domain~\cite{Martinis2003,Cywinski2008,Biercuk2010,Soare2014,Kabytayev2014,Alvarez2011}.
For Gaussian noise, decoherence is described by a function
\begin{equation}
\label{eq:W}	W(\tau) = \exp \left( -\int_0^\infty \frac{\drv f}{2 \pi^2} S(f) \frac{F(f \tau)}{f^2} \right),
\end{equation}
corresponding to a singlet probability $P_S(\tau)=\frac12[W(\tau) + 1]$.
In this expression, $F(f \tau)$ is a filter function that depends on the particular pulse sequence.

Filter functions for Hahn echo ($\FHahn$) and several CPMG sequences ($\FCPMG$) for fixed $\tau$ are plotted in Fig.~\ref{fig2}b (gray) for varying numbers of $\pi$ pulses.
We write the CPMG filter function as a product $\FCPMG = \tfrac{1}{2} \FFID \times \Fenv$, where $\FFID$ is the filter function corresponding to the free induction decay and $\Fenv$ is a slowly varying envelope (see Methods). $\Fenv$ is periodic with period $2/\tau$, with minima occurring at zero frequency and multiples of $2/\tau$ (Fig.~\ref{fig2}(b), black), independent of $n$.
Specific features of the filter functions can be exploited to decouple the qubit from its characterisitic noise environment.
First, for fixed separation time $T = n\tau$, the filter minimum near zero frequency becomes wider for increasing $n$ (i.e., decreasing $\tau$, note that the horizontal axis in Fig.~\ref{fig2}b is normalized frequency $f\tau$).
Thus for fixed $T$, decoupling from low frequency $1/f^\beta$-type noise ($\beta>0$) becomes more efficient as $n$ increases. 
Second, the minima that occur at multiples of $1/\tau$ indicate that noise at these frequencies  is notch-filtered, in the sense that specific narrow frequency windows are suppressed.

A schematic of the spectral density of nuclear noise for the $S$-$T_0$ qubit fabricated in a GaAs heterostructure is shown in Fig.~\ref{fig2}c, distinguishing longitudinal low frequency noise (green) and transverse narrow-band noise (red). 
The low frequency longitudinal contribution is well described by a power-law spectrum~\cite{Reilly2008,Medford2012}, and can be removed efficiently by any CPMG sequence (Fig.~\ref{fig2}d). 
The high frequency transverse contribution due to relative Larmor precession of nuclei is concentrated near the three Larmor frequency differences \cite{Cywinski2009a}, at megahertz frequencies for tesla-scale applied fields. Remarkably, two of the Larmor difference frequencies, $\fGb - \fGa$ and $\fGa - \fAs$, are nearly equal, independent of magnetic field, and hence the third frequency difference, $\fGb - \fAs$, occurs at twice that frequency. This coincidental property of the three nuclear species allows us to approximately align minima of the filter function with {\it all three} frequency differences  by correctly choosing the time between $\pi$ pulses, $\tau$, thereby decoupling the qubit from low and high frequency nuclear noise simultaneously.


\begin{figure}
	\includegraphics[scale=1]{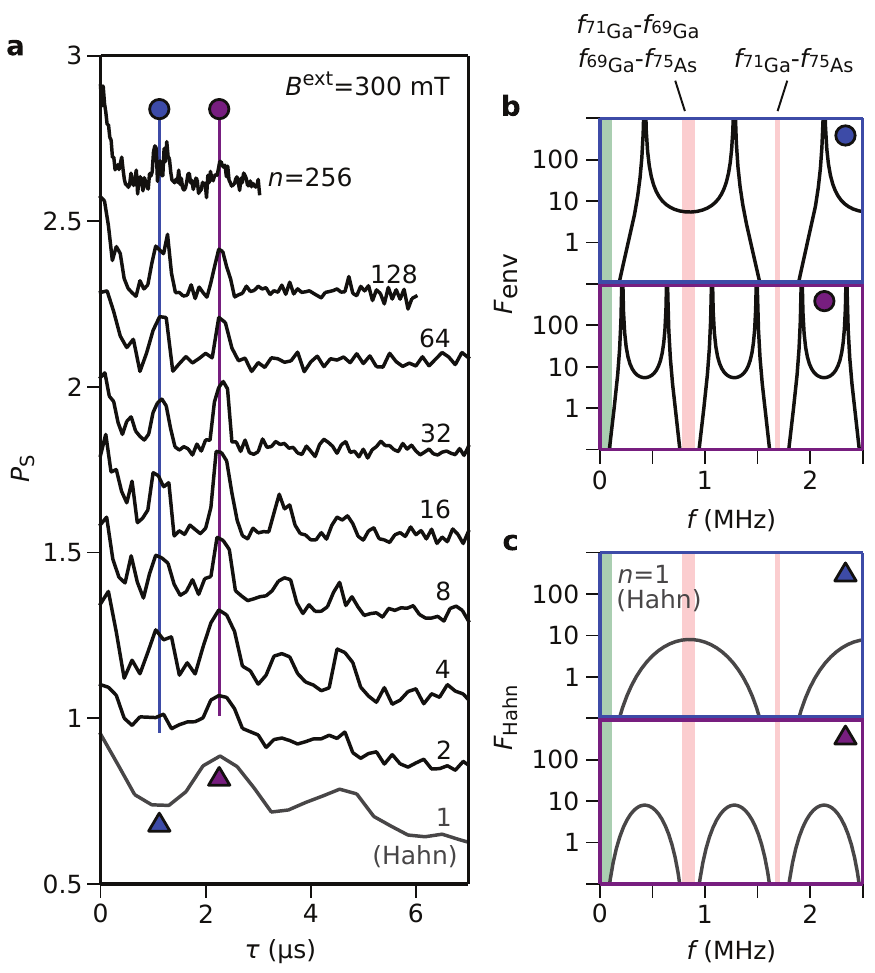}
	\caption{\textbf{Revival of coherence due to decoupling from nuclear Larmor precession.}
	\textbf{a}, Singlet return probability, $P_S$, as a function of the time between $\pi$ pulses, $\tau$, for various numbers of $\pi$ pulses, $n$. Curves are offset for clarity.
	\textbf{b}, Filter function envelope (black) and nuclear noise frequencies expected at 300 mT (shaded) for two choices of $\tau$. In both cases (marked by blue and purple lines in \textbf{a}) the revival in $P_S$ appears when minima of the filter function align with nuclear difference frequencies.  
	\textbf{c}, Filter function of Hahn-echo sequence for the same choices of $\tau$ as in \textbf{b}. The absence of the first revival (marked by a blue triangle in \textbf{a}) indicates that coherence is lost when the maximum of the filter function overlaps with the peaks in the nuclear noise spectrum (shaded). 
	The revival of $P_S$ for the second choice of $\tau$ (marked by the purple triangle in \textbf{a}) corroborates the destructive role of nuclear Larmor dynamics in qubit decoherence. 
	}
	\label{fig3}
\end{figure}

We now demonstrate the efficacy of this notch filter strategy in our experimental setup.
The narrow-band character of the high frequency nuclear noise is revealed by plotting the observed singlet return probability $P_S$ as a function of $\pi$-pulse separation time $\tau$ (rather than total separation time $T$).
Independent of the choice of $n$, we observe  an initial loss of coherence followed by revivals at $\tau\approx$ 1.1, 2.2, 3.3, ... $\mu$s (Fig.~\ref{fig3}a).
These values of $\tau$ correspond to decoupling conditions shown in Fig.~\ref{fig3}b, namely the alignment of nuclear difference frequencies (shaded red) with minima of the filter function envelope. 
Qualitatively, the alternating depth of filter minima in Fig.~\ref{fig3}b also explains the alternating heights of revivals, most pronounced for $n=4$ in Fig.~\ref{fig3}a. 
With increasing $\tau$, the height of the revivals decreases. This is related to decoherence arising from low frequency noise (shaded green in Fig.~\ref{fig3}b) \cite{Medford2012}.
Revivals observed for Hahn-echo sequences can be explained similarly, except that the filter function for $\tau\approx$ 1.1 $\mu$s has a maximum near  0.9 MHz (Fig.~\ref{fig3}c), rather than a minimum. Accordingly,  $P_S$ shows a minimum near $\tau=1.1$ $\mu\mathrm{s}$ instead of a revival (cf. $n=1$ data in Fig.~\ref{fig3}a).

\begin{figure*}
	\includegraphics[scale=1]{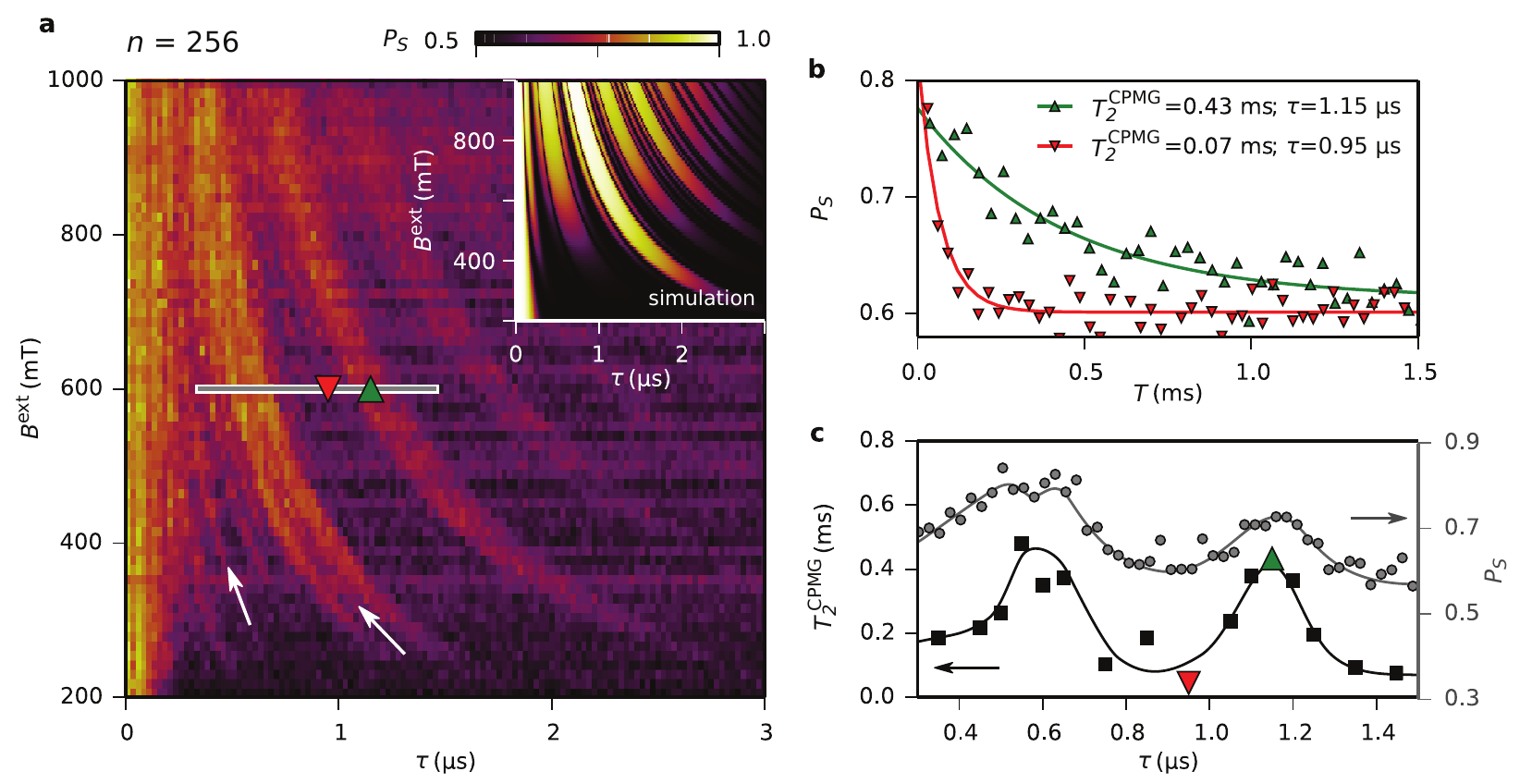}
	\caption{\textbf{Effect of magnetic field and $\mathbf{\tau}$ on qubit coherence.}
	\textbf{a}, Singlet return probability $P_S$ as a function of time between $\pi$ pulses, $\tau$, and external magnetic field, $\Bext$, for  a fixed number of $\pi$ pulses, $n=256$.
	Inset: A semiclassical model, generalizing the model of \cite{Neder2011} to the case of CPMG sequences, with no free parameters (see Supplementary Section 2). Arrows indicate fine features that the model fails to reproduce.
	\textbf{b}, Singlet return probability, $P_S$, measured as a function of total separation time $T=n\tau$, where $n$ is varied between 32 and 1536, for $\tau=0.95$ and $1.15$ $\mu$s at $\Bext=600$ mT, the values marked in \textbf{a}. Solid lines are fits to a decay law with exponential decay time $T_2^\mathrm{CPMG}$ (see Methods).
	\textbf{c}, Coherence time $T_2^\mathrm{CPMG}$, measured by increasing $n$ as in \textbf{b}, as a function of $\tau$ (squares). For comparison, $P_S$ for constant $n$, reproduced from the grey cut in \textbf{a}, is also shown (circles). Lines are guides to the eye. Triangles indicate the coherence times obtained from \textbf{b}. 
	}
	\label{fig4}
\end{figure*}

The dependence of the decoupling condition for $\tau$ on nuclear Larmor dynamics can be verified by changing the applied magnetic field. 
In Fig.~\ref{fig4}a we fix $n=256$ and measure the decay of coherence as a function of $\Bext$.
As expected for a linear nuclear Zeeman splitting we find that the positions of the revival peaks  follow a $1/\Bext$ dependence.
We further observe that the peaks in $P_S(\tau)$ disappear at low magnetic fields.
This may arise from several effects.
First, the transverse Overhauser field in each dot, $\vec{B}^{{\rm nuc}}_{\perp}$, affects the total electronic Zeeman energy more strongly at low magnetic field (see Fig.~\ref{fig1}d), thereby accelerating dephasing.
Second, the energy mismatch between nuclear and electron Zeeman splittings becomes smaller at low fields, increasing electron-mediated interactions between nuclear spins and the associated low frequency noise \cite{Yao2006,Deng2005,Cywinski2009a}.
Third, an increase in $\tau$, as needed to maintain the decoupling condition at lower fields, narrows the filter function minima and thus reduces decoupling from high frequency noise. 

Next we show that revivals in $P_S$ translate to prolonged qubit coherence times, by increasing $n$ while keeping $\tau$ and $\Bext$ fixed. 
This method, pioneered in NMR~\cite{Carr1954}, differs from other spin qubit experiments in which $n$ is held constant while $\tau$ is swept proportionally to $T$~\cite{Bluhm2011,Medford2012}.
Figure~\ref{fig4}b plots decay curves $P_S(T=n\tau)$  obtained for $\tau=0.95$ and $1.15$ $\mu$s at $\Bext=600$ mT (the corresponding points are indicated in Fig.~\ref{fig4}a). 
For large $n$ and Gaussian noise, an exponential decay of coherence is expected, independent of the power spectrum of the noise~\cite{Alvarez2011}. 
By fitting exponential decay curves~\cite{Carr1954,Alvarez2011} (see Methods) we extract drastically different coherence times $\TCPMG$, as indicated. 
Values of $\TCPMG$ for more choices of $\tau$ are plotted in Fig.~\ref{fig4}c, along with $P_S(\tau)$ extracted from Fig.~\ref{fig4}a. 
We observe a clear correlation between $\TCPMG$ and $P_S(\tau)$, indicating that qubit coherence is significantly prolonged whenever the decoupling condition is fulfilled. 
The exponential decay indicates that coherence is limited by either incompletely filtered longitudinal noise or pulse errors.

\begin{figure}
	\includegraphics[scale=1]{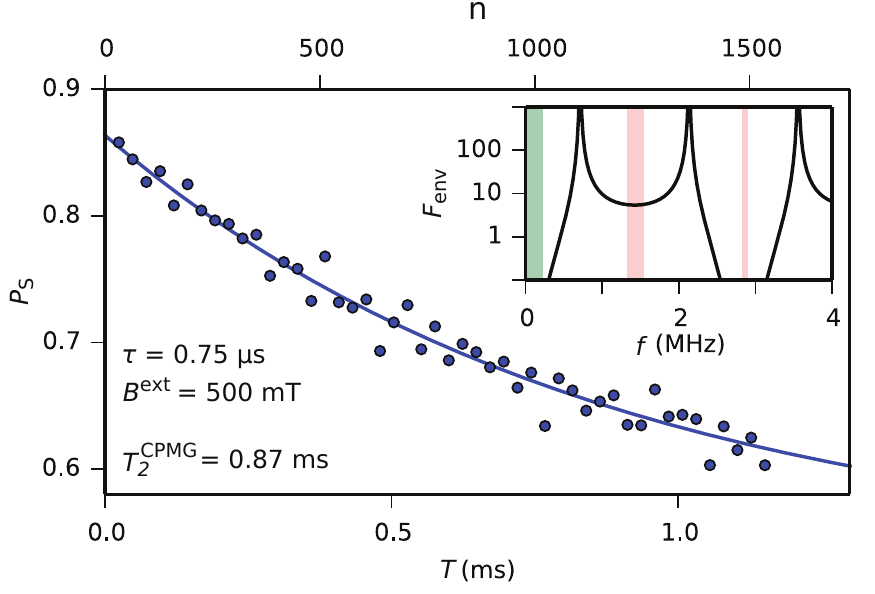}
	\caption{
	\textbf{Singlet return probability, $\mathbf{P_S}$, as a function of total separation time, $\mathbf{T=n\tau}$, for optimized and fixed values of $\mathbf{\tau}$ and $\mathbf{\Bext}$.} 
	An exponential fit to the data yields $T_2^\mathrm{CPMG} = 0.87 \pm 0.13$ ms. The data points correspond to the CPMG sequences with $n=32$ to $n=1536$ $\pi$ pulses. Inset: Alignment of the filter function envelope with nuclear noise spectrum for this choice of $\tau$ and $\Bext$.
	}
	\label{fig5}
\end{figure}

Finally we comment on the limits of preserving qubit coherence. 
Most of the observed features in Fig.~\ref{fig4}a are captured by a generalization of the semiclassical model of Ref.~\cite{Neder2011}, modified to include the details of the CPMG pulse sequence.
The model involves four device-specific parameters (Fig.~\ref{fig4} inset): 
the effective number of nuclei interacting with each electron, $N=7\times10^5$, 
a phenomenological broadening, $\delta B=1.1$ mT,  of the effective magnetic field acting on nuclei (likely due to quadrupolar splitting arising from electric field gradients \cite{Bluhm2011,Neder2011,Botzem2016}),
the spectral diffusion time, $T_\mathrm{SD} = 600$ $\mu$s, 
and the exponent associated with the linear low frequency noise, $\beta=3$ (all determined by independent measurements as described in Supplementary Sections 3 and 4). 
The model suggests that the longest coherence time may be achieved by choosing a decoupling condition corresponding to the second revival at high magnetic fields, consistent with a reduced contribution of $\vec{B}^\mathrm{nuc}_{\perp}$ to $\Btot$ in each dot (see Fig.~\ref{fig1}d) and the decoupling condition depicted in Fig.~\ref{fig2}d. 
We note that the model does not take pulse errors into account and does not show several fine features observed in experiment (see white arrows in Fig.~\ref{fig4}a, see Supplementary Section 5). 

By exploring the parameter space between $\Bext=300$ and $1000$ mT with $\tau$ corresponding to the first revival peak, we observe coherence times around 0.7 ms for $\Bext=500$ to 600~mT, with the largest being $\TCPMG = 0.87 \pm 0.13$ ms (Fig.~\ref{fig5}), measured at $\Bext=500$ mT and $\tau=0.75$ $\mu$s. However, the number of examined values of $\Bext$ and $\tau$ remains insufficient to resolve the fine structure apparent in the first revival peak.

We expect further improvements by using shorter $\pi$ pulses and nuclear programming \cite{Foletti2009}. This will improve the fidelity of $\pi$ pulses and suppress low frequency noise, allowing the advantageous use of the decoupling condition in Fig.~\ref{fig2}d at high magnetic fields and high pulse rates. 


In summary, dynamical decoupling sequences were demonstrated to provide decoupling from narrow-band high frequency noise, acting as a notch filter for the nuclear environment. This technique was used to efficiently decouple a GaAs-based $S$-$T_0$ qubit from its nuclear environment. By synchronizing the repetition rate of $\pi$ pulses in CPMG sequences with differences of nuclear Larmor frequencies, the coherence of a $S$-$T_0$ qubit coupled to nuclear spin bath was extended the millisecond regime (0.87 ms), five orders of magnitude longer than the gate operation time.

\section{Acknowledgements}
We thank Rasmus Eriksen for help in preparation of the reflectometry setup.
This work was supported by IARPA-MQCO, LPS-MPO-CMTC, the Polish National Science Centre (NCN) under Grant No.~DEC-2012/07/B/ST3/03616, the Army Research Office, the Villum Foundation and the Danish National Research Foundation. 

\section{Author contributions}
S.F., G.C.G. and M.J.M. grew the heterostructure. P.D.N. fabricated the device. F.M., P.D.N., F.K. and F.K.M. prepared the experimental setup. F.K.M., F.M. and F.K. performed the experiment. E.B., \L.C. and M.S.R. developed theoretical model and performed simulations. F.K.M., F.K., F.M., E.B., \L.C., M.S.R. and C.M.M. analysed data and prepared the manuscript.

\section{Additional information}
Supplementary information is available in the online version of the paper. Reprints and permission information is available online at www.nature.com/reprints. Correspondence and requests for materials should be addressed to F.K. 

\section{Competing financial interests}
The authors declare no competing financial interests

\bibliographystyle{naturemag}
\bibliography{bibliography}{}

\section{Methods}

\subsection{The sample}

The sample, identical to the one shown in Fig.~1a, is fabricated from a GaAs/AlGaAs quantum well grown by molecular beam epitaxy. 
Crystallographic axes are shown in Fig.~1a. 
A high-mobility 2D electron gas (2DEG) is formed 57 nm below the sample surface with carrier density $n_s$~=~$2.5\times10^{15}$~m$^{-2}$ and mobility $\mu$~=~230~m$^{2}$/Vs. Metallic gates, separated from the heterostructure by a 10 nm layer of HfO$_2$, are used to confine two electrons in the region indicated by a rectangle in Fig.~1a.  Gates indicated in blue and red are operated at negative voltages to deplete the 2DEG underneath, while gates colored in green are biased with positive voltages to accumulate electrons beneath. The charge state and tunnel coupling of the double dot can be controlled on a nanosecond timescale by applying voltage pulses to gates $\VL$, $\VM$, and $\VR$. 

\subsection{Initialization and readout of the qubit}

The sample is measured at a base temperature of 25~mK in a cryofree dilution refrigerator, with an external magnetic field $\Bext$ applied parallel to the $z$ direction indicated in Fig.~1a. 
The qubit is initialized in a singlet state by tilting its charge state into the (2,0) charge configuration and allowing the exchange of electrons with the left lead near the (1,0) charge transition \cite{Petta2005}.

After qubit manipulation the state of the qubit is measured by tilting the double well potential to favour the (2,0) charge state. If the two electrons are in the spin triplet configuration, Pauli blockade prevents reaching the (2,0) state, and the charge configuration remains (1,1).
The charge state of the double dot modifies the conductance through a proximity sensor dot operated as a single electron transistor. This sensor dot is embedded in a radio-frequency resonant circuit, enabling us to distinguish singlet and triplet states in 8 $\mu$s with a readout visibility of approximately 80\%, as defined in Ref. \cite{Barthel2010}.

\subsection{ Envelope of a filter function for CPMG sequence } 

Filter functions for Hahn echo and CPMG sequence (for even number of $\pi$ pulses, $n$) are given by \cite{Cywinski2008}
\begin{equation}
	\FHahn(f \tau) = 8 \sin^4 \left( \pi f \tau/2 \right);
\end{equation}

\begin{equation}
	\FCPMG(f \tau) = \frac{ 8 \sin^4 \left( \pi f \tau / 2 \right) \sin^2 \left(\pi f \tau n \right) }{ \cos^2 \left(\pi f \tau \right) }.
\end{equation}
\vspace{5pt}

To emphasize the qualitative difference between CPMG sequences and the Hahn echo sequence, and represent features of CPMG filter functions relevant for large number of $\pi$ pulses, $n$, we rewrite
\begin{equation}
	\FCPMG = \frac{1}{2} \Fenv \times \FFID
\end{equation}
using $n$-independent filter function envelope
\begin{equation}
	\Fenv(f \tau) = \frac{ 8 \sin^4 \left( \pi f \tau / 2 \right) }{ \cos^2 \left(\pi f \tau \right) }. 
\end{equation}
obtained by dividing
 $\FCPMG$ by the filter function corresponding to free induction decay
\begin{equation}
	\FFID(f T) = 2 \sin^2(\pi f T).
\end{equation}
Here $T=\tau n$ corresponds to a free induction decay time equal to the total duration as a CPMG sequence. This normalization removes a fine comb related to the total length of the sequence.

\subsection{Exponential fits to $P_S(T)$}

In contrast to many spin qubit experiments~\cite{Bluhm2011,Petta2005,Dial2013,Botzem2016,Barthel2010,
Medford2012,Cywinski2009,Cywinski2009a,Kawakami2016,DeSousa2003,Yao2006,Childress2006}
we measure coherence not by keeping $n$ constant and sweeping $\tau$, but by increasing $n$ while keeping $\tau$ constant. This method, which is standard in NMR experiments~\cite{Carr1954}, results in an exponential decay of coherence for large number of $\pi$ pulses $n$ and long evolution times $T=n\tau$, independent of the power spectrum of the Gaussian noise~\cite{Alvarez2011}. The rate of such a decay is determined by the noise spectrum at a frequency corresponding to the first peak of the filter function from Fig.~\ref{fig2}d at $f=1/2\tau$.

Therefore, we perform an exponential fit of the form $A+B\exp(-T/\TCPMG)$ to the data, where $A$ and $B$ account for preparation and readout fidelity as well as rapid initial decay of the signal~\cite{Bluhm2011,Botzem2016}, and $\TCPMG$ is a coherence time of the qubit. Typical values of $A$ and $B$ obtained from fits used to extract values of $\TCPMG$, shown in Fig.~\ref{fig4}c, are $A\sim 0.6$ and $B\sim 0.2$. Fit to the data presented in Fig.~5 yields $A=0.53$ and $B=0.34$. 

\onecolumngrid

\newpage

\renewcommand{\thefigure}{S\arabic{figure}}  
\renewcommand{\theequation}{S\arabic{equation}}
\renewcommand{\thetable}{S\arabic{table}}
\setcounter{figure}{0}   
\setcounter{equation}{0}   

\renewcommand{\theHfigure}{S.\thefigure}

\begin{center}
	\large \bf Spectrum of the GaAs nuclear environment -- supplementary information
\end{center}

\begin{center}

Filip K. Malinowski, Frederico Martins, Peter D. Nissen, Edwin Barnes, \L{}ukasz Cywi\'nski, Mark S. Rudner, Saeed Fallahi, Geoffrey C. Gardner,
Michael J. Manfra, Charles M. Marcus, and Ferdinand Kuemmeth

\vspace{0.5cm}

\begin{minipage}{0.8\textwidth}
	\small
	The supplementary information is divided into sections, which discuss the following topics:
	\begin{itemize}
		\item[1.] Calibration of $\pi$ pulses
		\item[2.] Semiclassical model of decoherence due to nuclear noise
		\item[3.] Estimating $N$ and $\delta B$ from Hahn echo signal
		\item[4.] Estimating $T_\mathrm{SD}$ and $\beta$ from scaling of coherence time
		\item[5.] Extension of the model to take into account anisotropy of electron $g$-factor. Discussion of the origin of the splitting of the first revival peak.
	\end{itemize}
\end{minipage}
\end{center}

\twocolumngrid

\subsection{1. Calibration of $\pi$ pulses}

To generate decoupling sequences consisting of as many as 1000 $\pi$ pulses we took advantage of charge-noise-insensitive symmetric exchange pulses. This new technique improves the quality factor of exchange oscillations by a factor of six relative to conventional method of tilting the double dot potential~\cite{Martins2016,Reed2016}. A detailed analysis of this technique, and results obtained in the preceding experiment from the same setup and same sample, can be found in Ref.~\cite{Martins2016,Barnes2016}.

\begin{figure}
	\includegraphics[scale=1]{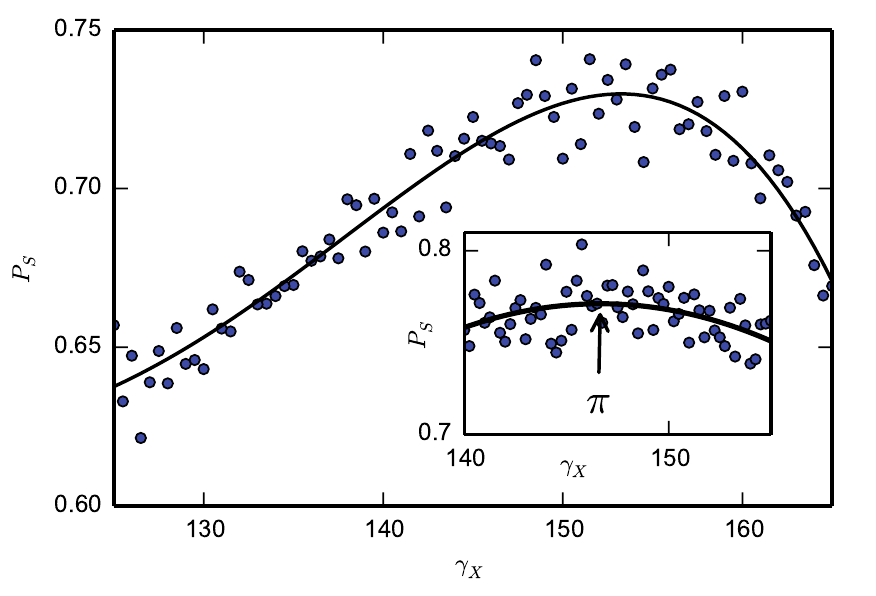}
	\caption{
	\textbf{Calibration of $\pi$ pulses} Singlet return probability $P_S$ as a function of a symmetric exchange pulse amplitude, for exchange time of 4.167 ns in a Hahn echo experiment. The maximum probability indicates $\pi$ pulse. Solid line is a guide to the eye. Inset: around the maximum the parabola is fitted to the data. Symbol $\pi$ indicates value of $\gamma_X$ corresponding to the $\pi$ pulse.
	}
	\label{figS1}
\end{figure}

The optimization was performed by maximizing the Hahn echo signal by varying the amplitude of the symmetric exchange pulse, $\gamma_X$, while keeping detuning $\varepsilon_X=0$ mV, exchange time $t_X=4.167$ ns and total evolution time $\tau=0.75$ $\mu$s fixed (Fig.~\ref{figS1}). The experiment was performed on the same device and in identical tuning as Ref.~\cite{Martins2016}, where $\gamma_X$ and $\varepsilon_X$ are defined and discussed in detail.

We note that symmetric exchange pulses show a weaker dependence on gate voltages than ordinary tilt pulses.
Hence, symmetric $\pi$ pulses are more robust against fluctuations of pulse amplitudes. 
On the other hand, symmetric pulses require a larger amplitude, resulting in somewhat slower exchange gates compared to conventional tilted exchange gates. This limitation makes $\pi$ pulses more susceptible to errors induced by gradients of the Overhauser field, causing a tilted rotation axis of the qubit. In future experiments, larger pulse amplitudes can be achieved straightforwardly by decreasing the attenuation in the transmission lines in the cryostat.

Nevertheless, 
the CPMG sequence is particularly robust to
two kinds of $\pi$ pulse errors that affect exchange gates~\cite{Borneman2010}. The first is over or under rotation around the vertical axis of the Bloch sphere due to miscalibration of pulse amplitude and duration. 
The second is tilt of the rotation axis in the $\ket{S}$-$\ket{T_0}$--$\ket{\ud}$-$\ket{\du}$ plane due to uncontrolled gradients of the Overhauser field. 

\subsection{2. Semiclassical model for decoherence}
The inset of Fig.~4 shows theoretical results for coherence revivals.
The model is derived closely following the semiclassical approach developed in Ref.~\cite{Neder2011}.
The starting point is to express the Hamiltonian for the $\ket{\ud}$, $\ket{\du}$ subspace of the two-spin system as
\begin{equation}
\label{eq:NoiseHam}	\hat{H}(t) = g^* \mu_B \sum\limits_{d=L,R} \left( B_{z,d}^{\rm nuc}(t) + \frac{|\mathbf{B}_{\perp,d}^{\rm nuc}(t)|^2}{2|\mathbf{B}^{\rm ext}|}\right)  c(t) \hat{S}_d^z,
\end{equation}
where $g^*$ is the electron $g$-factor, $\mu_B$ is the Bohr magneton, $\mathbf{B}^{\rm ext}$ is the external magnetic field, $B_{z,d}^{\rm nuc}$  ($\mathbf{B}_{\perp,d}^{\rm nuc}$) is the Overhauser field component parallel (perpendicular) to external magnetic field, $\hat{S}_d^z$ is the electron spin operator, $d=L,R$ labels the left and right dots, and we assume $|\mathbf{B}^{\rm ext}| \gg |\mathbf{B}_{d}^{\rm nuc}|$.
Here, the sequence of $\pi$ pulses applied to the qubit is captured by
\begin{equation}
	c(t) = \sum\limits_{j=0}^n (-1)^j \theta(t_{j+1}-t) \theta(t-t_j),
\end{equation}
where $t_j$ is the time at which the $j$-th $\pi$ pulse of the CPMG sequence is applied (with $t_0=0$, $t_{n+1}=T$), and $\theta(t)$ is the Heaviside step function. 

Reference \cite{Neder2011} treated only the Hahn echo sequence.
This corresponds to $n = 1$ in the above.
Following the same sequence of steps, we obtain results for arbitrary $n$.
As in that case, the decoherence function $W(\tau) = W_z(\tau)W_\perp(\tau)$ is separated into a product of contributions from the longitudinal and transverse noise sources.
The low-frequency longitudinal noise contribution is of the form $W_z(\tau) = e^{-(\tau/T_\mathrm{SD})^{\alpha}}$, where $\alpha=\beta+1$ is related to the exponent in the power law $1/f^\beta$ describing the spectrum of this noise source~\cite{Medford2012}, and $T_\mathrm{SD}$ is the spectral diffusion time.
Because the transverse field enters the Hamiltonian (\ref{eq:NoiseHam}) as a square, $|\mathbf{B}_{\perp,d}^{\rm nuc}(t)|^2$, this noise source is effectively non-Gaussian \cite{Cywinski2014,Neder2011}.
As a result, the decoherence function for dot $d$ is of the form $W_{\perp,d}(\tau) = 1/\det(1 + iT_d)$ with components of the matrix $T_d$ given by
\begin{widetext}
\begin{equation}
	T_{kl,d} = \frac{5 A_{\xi(k)} A_{\xi(l)} \sqrt{N_k N_l}}{2 g^* \mu_B |\mathbf{B}^{\rm ext}|}
	\frac{\omega_{kl}}{\omega_{kl}^2 - A_{kl}^2}
	\left\lbrace 1- \frac{\cos\left( \frac{A_{kl} T}{2n} \right)}{\cos \left( \frac{\omega_{kl} T}{2n} \right)} \right\rbrace
	\sin\left( \frac{\omega_{kl} T + n\pi}{2} \right) e^{i \frac{\omega_{kl}T + n\pi}{2}}.
\end{equation}
\end{widetext}
Here $k,l$ labels groups of nuclei associated by isotope and local nuclear Zeeman coupling, $A_{\xi(k)}$ is the hyperfine coupling constant for nuclei in group $k$, $A_{kl}=A_{\xi(k)}-A_{\xi(l)}$, $N_k$ is number of nuclei in a group, $\omega_{kl} = \omega_k - \omega_l$ is a difference of Larmor frequencies between nuclei from two groups, and $T=n\tau$ is the total evolution time.
Specifically, the nuclei of each isotope are divided into $K$ groups using the relation $N_k=n_{\xi(k)}N/(2K)$, where $n_\xi$ is the number of nuclei of isotope $\xi$ per unit cell, and where all nuclei within each group have the same Larmor angular frequency $\omega_k$.  
The value of $\omega_k$ for each group is drawn from a Gaussian distribution centered at the bare Larmor frequency $\omega_\xi$ for the corresponding isotope, with standard deviation $\delta B$. 
The broadening $\delta B$ is introduced as a phenomenological parameter to take into account an effective spread in the Larmor frequencies due to inhomogeneous quadrupolar splittings and dipole-dipole interactions.
For the simulation shown in Fig.~4, differences between hyperfine couplings within the same isotope were neglected, and convergence was obtained with $K=4$ groups.

Larmor frequencies and hyperfine couplings used to perform the simulation shown in Fig.~4 were taken from Ref.~\cite{Cywinski2009a} (table \ref{tab1}). The remaining parameters are: the effective number, $N$, of nuclei interacting with each electron, inhomogeneity, $\delta B$, of the effective magnetic field acting on the nuclei, the spectral diffusion time, $T_\mathrm{SD}$, and the exponent, $\beta$, associated with the low-frequency noise. The following sections explain how these parameters are obtained.

\begin{table}[b]
	\caption{Bare Larmor angular frequencies, $\omega_\xi$, hyperfine constants, $A_\xi$, in units of angular frequency, and abundances, $n_\xi$, of $^{69}$Ga, $^{71}$Ga and $^{75}$As, taken from Ref. \cite{Cywinski2009a}.}
	\renewcommand\tabcolsep{10pt}
	\begin{tabular}{c|ccc}
		& $\omega_\xi/B$ [s$^{-1}$T$^{-1}$] & $A_\xi$ [s$^{-1}$] & $n_\xi$\\ \hline
		$^{69}$Ga & 64.2 & $5.47\times10^{10}$ & 0.604\\
		$^{71}$Ga & 81.6 & $6.99\times10^{10}$ & 0.396\\
		$^{75}$As & 45.8 & $6.53\times10^{10}$ & 1\\
	\end{tabular}
	\label{tab1}
\end{table}

\subsection{3. Estimating $N$ and $\delta B$ from Hahn echo signal}

\begin{figure}
	\includegraphics[scale=1]{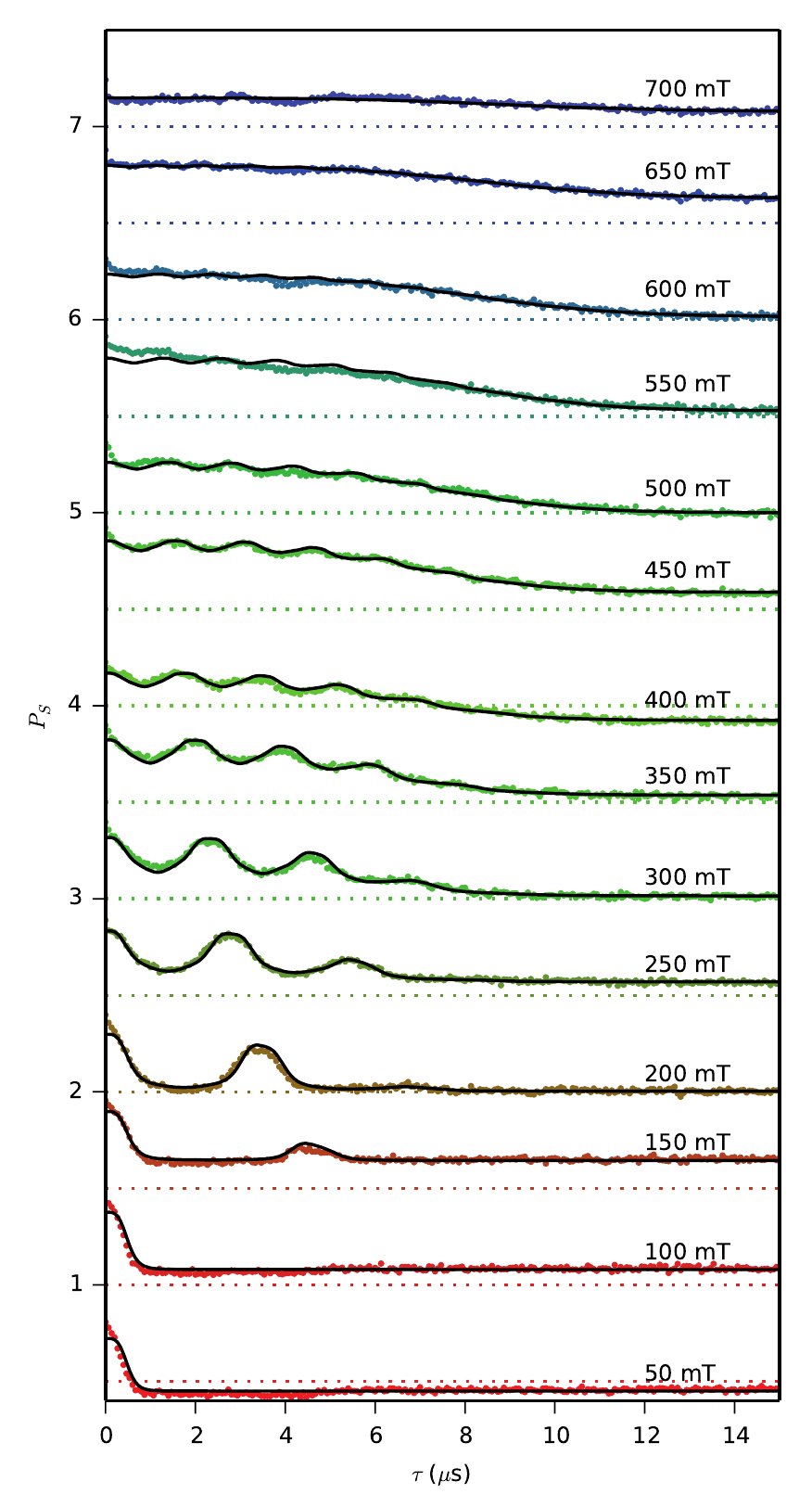}
	\caption{\textbf{Revival of coherence under Hahn-echo sequence.} Singlet return probability $P_S$ as a function of separation time $\tau$ for various magnetic fields. Datasets are offset for clarity. Dotted lines indicate $P_S=0.5$ for data plotted in corresponding color. Black lines are simulations with $\delta B = 1.1$ mT, $T_\mathrm{SD}^\mathrm{Hahn} = \infty$, $N = 7\times 10^5$. They are fitted to experimental data using offset and visibility, different for each curve.}
	\label{figS2}
\end{figure}

The simulation of revivals under CPMG sequences requires knowledge of four device-specific parameters, two of which, the effective number, $N$, of nuclei interacting with each electron and the inhomogeneity, $\delta B$, of the effective magnetic field acting on the nuclei, are extracted from Hahn echo data obtained at several magnetic fields (Fig.~\ref{figS2}). Following previous work \cite{Bluhm2011,Neder2011} we first fit theory to Hahn echo data at each magnetic field separately, keeping $\delta B$, $N$, the spectral diffusion time for Hahn echo, $T_\mathrm{SD}^\mathrm{Hahn}$, vertical offset, and vertical scaling as free parameters.
Setting $T_\mathrm{SD}^\mathrm{Hahn} \gg 1$ ms gives essentially equally good fits (i.e. $T_\mathrm{SD}^\mathrm{Hahn}$ cannot be determined accurately by this method) but values for $\delta B$ and $N$ obtained at various magnetic fields (150-350 mT) differ from each other by less than 20\%. Therefore we average these values and obtain  $\delta B = 1.1$ mT and $N = 7\times 10^5$. Fixing these values and $T_\mathrm{SD}^\mathrm{Hahn}=\infty$, we leave the vertical offset and vertical scaling as the only free parameters, and obtain excellent agreement for all magnetic fields, as seen in Fig.~\ref{figS2}. Visibility and offset are left as free parameters, independent for each curve, to accommodate a fluctuating readout visibility that is likely due to a buildup of the gradient of Overhauser field for large $\Bext$ \cite{Barthel2012}.

The only systematic deviation between the experimental results and the model is a slight, rapid, initial decay of the signal (first 3-5 data points of each data set). This effect was also observed in Refs.~\cite{Bluhm2011,Botzem2016}. 
The effect depended on the external magnetic field as well as the gradient of the Overhauser field~\cite{Botzem2016}, and was speculated to be related to the entanglement of the qubit with the nuclear bath or to $\pi$ pulse errors~\cite{Bluhm2011}.

\subsection{4. Estimating $T_\mathrm{SD}$ and $\beta$ scaling of coherence time}

To estimate the spectral diffusion time $T_\mathrm{SD}$ for the simulation in Fig.~4a we quantify the scaling of the CPMG coherence time with the number of $\pi$ pulses $n$ in a regime where revivals are not yet developed (i.e., for $n\leq 32$ at 750 mT)  \cite{Medford2012}.  
The coherence time is found to be proportional to $n^\gamma$, with $\gamma \sim 0.75$ \cite{Malinowski2016}.
Using this scaling behaviour we infer $T_\mathrm{SD} \approx 0.6$ ms for a CPMG sequence with 256 $\pi$-pulses. Using the relationship $\beta = \gamma / (1-\gamma)$ \cite{Medford2012} the exponent $\gamma=0.75$ corresponds to a power law of low-frequency noise governed by $1/f^\beta$ behaviour, with $\beta=3$, in reasonable agreement with previous measurements \cite{Medford2012}.

\subsection{5. Splitting of the first revival peak}

A possible explanation for the observed splitting in the first revival peak (Fig.~4a) is based on the anisotropy of the electronic $g$-factor. The $g$-factor anisotropy between [011] and [01-1] primary axes can be as high as 15\% in asymmetric GaAs/AlGaAs quantum wells \cite{Nefyodov2011}. In Ref.~\cite{Botzem2016} it was shown that the anisotropy has a strong impact on $S$-$T_0$ qubit coherence when the magnetic field is not parallel to one of the main axes. The combination of the anisotropy and small misalignment of the external magnetic field with the [011] crystal axis changes the magnetic field term in the system Hamiltonian (\ref{eq:NoiseHam}) to:
\begin{equation}
\label{eq:BAniso}
B_{z,d}^{\rm nuc}(t) + \frac{|\mathbf{B}_{\perp,d}^{\rm nuc}(t)|^2}{2|\mathbf{B}^{\rm ext}|} + \frac{g_\perp}{g_\parallel} \left[ B_{x,d}^{\rm nuc}(t) + B_{y,d}^{\rm nuc}(t) \right],
\end{equation}
where $g_\parallel$ ($g_\perp$) are diagonal (off-diagonal) elements of a $g$-tensor in the basis set by direction of the external magnetic field. The latter leads to the appearance of individual nuclear Larmor frequencies  in the nuclear noise spectrum in addition to nuclear difference frequencies \cite{Botzem2016}. As a result, the CPMG sequence will not be as efficient in suppressing nuclear noise even when the pulses are commensurate with all three difference frequencies.

\begin{figure*}
	\includegraphics[scale=1]{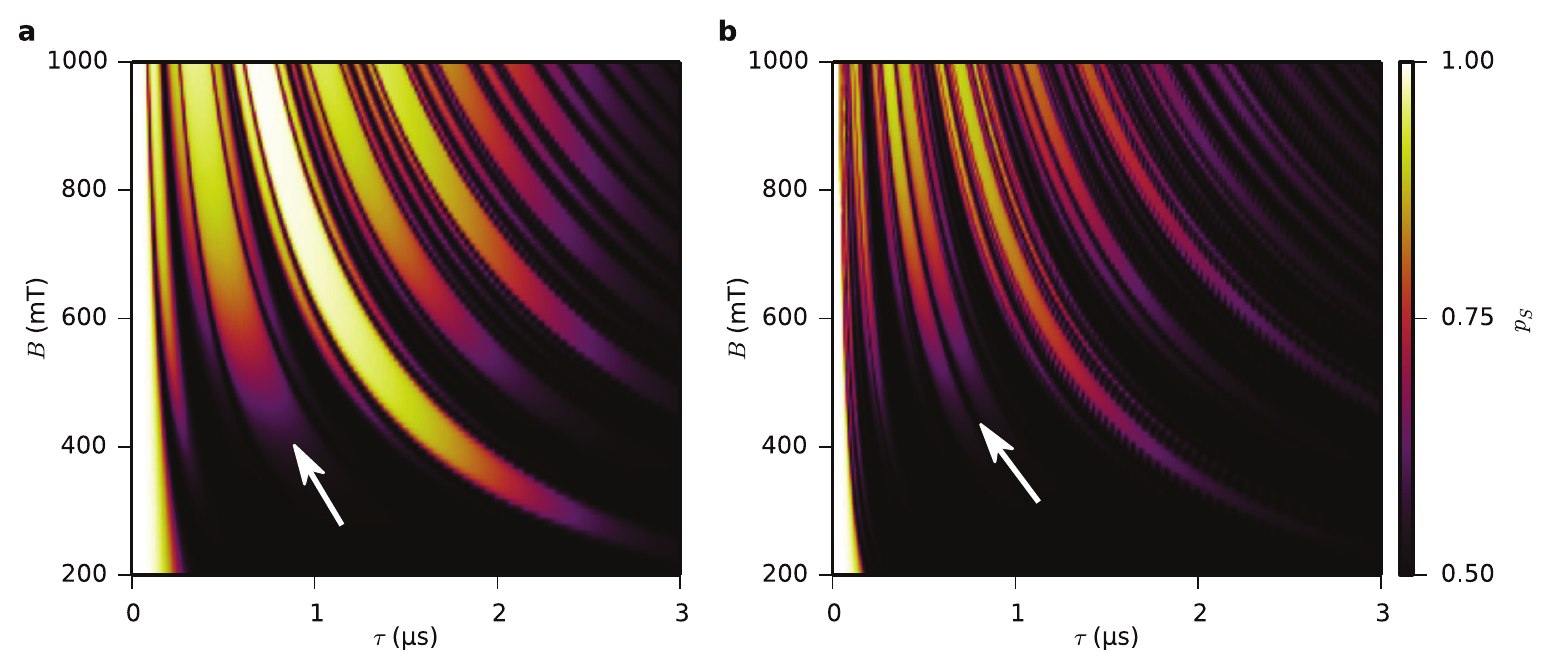}
	\caption{\textbf{Simulation of revivals of coherence under CPMG sequence for 256 $\mathbf{\pi}$ pulses.} \textbf{a}, Simulation omitting effects of g-factor anisotropy, identical to map in the inset of Fig.~4a, i.e., $g_\perp / g_\parallel = 0$. \textbf{b}, Simulation assuming $g_\perp / g_\parallel= 0.01$.}
	\label{figS3}
\end{figure*}

In Fig.~\ref{figS3} we present simulations showing the consequences of $g$-factor anisotropy. Panel (a) shows the simulation presented in the inset of Fig.~4a, i.e. $g_\perp / g_\parallel = 0$. In panel (b) we show a simulation that assumes $g_\perp / g_\parallel = 0.01$. Although our external magnetic field was nominally aligned with the [011] crystal axis (cf. Fig. 1a), the choice of $g_\perp / g_\parallel = 0.01$ is consistent with the smallest value observed in ~\cite{Botzem2016} for the same direction of magnetic field as in our setup. 
In our simulation a splitting of the first revival peak appears (indicated by a white arrow) as well as more complex fine structure in other revival peaks. Such fine structure is beyond the resolution of the experimental data.

We note that a splitting of the first revival peak appears exactly when the frequency of $\pi$ pulses coincides with a difference of Larmor frequencies $\fGb - \fGa$ and $\fGa - \fAs$. Therefore other mechanisms might also lead to the appearance of the splitting. We speculate that weak driving of the nuclei by a periodic Knight field could enhance flip-flops between nuclei of different species and therefore increase spin diffusion, leading to faster decoherence.

\onecolumngrid

\end{document}